\documentclass{ws-ijgmmp}

\usepackage{comment}

\def\ba{\begin{eqnarray}}
\def\ea{\end{eqnarray}}

\begin{document}


%
\catchline{}{}{}{}{}
%

\title{A modified gravity model coupled to a Dirac field in 2D spacetimes with quadratic nonmetricity and curvature}

\author{Caglar Pala$^{1}$\footnote{caglar.pala@gmail.com}\;\;, Ertan Kok$^{2}$\footnote{ekok@pau.edu.tr}\;\;, Ozcan Sert$^{2}$\footnote{osert@pau.edu.tr}\;\;, Muzaffer Adak$^{2}$\footnote{madak@pau.edu.tr}
}
\address{$^1$ Department of Physics, Erciyes University, 38280 Kayseri, Turkey \\
$^2$ Department of Physics, Pamukkale University, 20017 Denizli, Turkey}

\maketitle

\begin{history}
\end{history}

\begin{abstract}
After summarizing basic concepts for the exterior algebra we firstly discuss the gauge structure of the bundle over base manifold for deciding the form of the gravitational sector of the total Lagrangian in any dimensions. Then we couple minimally a Dirac spinor field to our gravitational Lagrangian 2-form which is quadratic in the nonmetricity and both linear and quadratic in the curvature in two dimensions. Subsequently we obtain field equations by varying the total Lagrangian with respect to the independent variables. Finally we find some classes of solutions of the vacuum theory and then a solution of the Dirac equation in a specific background and analyse them.
\end{abstract}


\section{Introduction}

Although the Einstein's theory of gravity is mathematically elegant and very successful in the solar system, there are strong reasons for modifications motivated by recent astrophysical and cosmological observations and by vain efforts on its quantization. That can be accomplished in various ways. For example, several works attempt to modify the Einstein-Hilbert Lagrangian by adding higher order curvature scalar terms \cite{Kretschmann1917}, \cite{Weyl1919} which later on standing for the modern $f(R)$ theories \cite{Bergmann1968}, \cite{Buchdahl1970}, \cite{Starobinsky1980} and also there are models including torsion \cite{Hammond2002}, \cite{Dereli2013} and nonmetricity \cite{Adak2006_1}, \cite{demir2012} into the geometry.    Especially in the metric-affine gravity (MAG) models, that separate the metric and the affine connection as independent variables, the gauge approach is extensively used \cite{Tomi2018}. However, such these models possess Cauchy problems that still an open area \cite{Nicolas2007}, \cite{Hehl1999}. The researchers of a very recent paper \cite{jimenez2020} discuss the presence of ghostly instabilities in detail for metric-affine theories constructed with higher order curvature terms. In addition, as an alternative approach through the path, Kaluza and Klein studied first the viable unification of gravity with electromagnetism within pure geometric context by using Einstein-Hilbert term only, but in addition proposing extra dimension to spacetime \cite{KaluzaKlein}. This idea soon has considerably attracted attention and nowadays find place in modern string-type theories. In order to implement the quantization methods -which being a solid underlying philosophy regarding to the Standart Model- to the gravity one might study on De Witt's perturbative \cite{Lyra1992} and Witten's non-perturbative approaches \cite{Witten1988}, \cite{Deser1989}.

On the other hand the gauge theory is a successful approach in the modern physics for explaining the nature. For example, the electromagnetic and the weak nuclear interactions have been combined through $SU(2) \otimes U(1)$-gauge theory; the electroweak theory. $SU(3) \otimes SU(2) \otimes U(1)$-gauge theory which is one step advance of the same approach contains strong nuclear interactions as well; the so-called standard model \cite{griffiths1987}. In this context the author of the reference \cite{adak2018} states that gravity can be formulated by a locally gauge invariant Lagrangian in which the metric represents the gravitational field and the full connection is interpreted as the gauge potential in the theory of symmetric teleparallel gravity. In this model one must constrain the non-Rimennian curvature to vanish. When one looks at the analogy to the Maxwell-Dirac theory, the zero-curvature constraint may be seen excessive restraint, see the table \ref{tab:analogy}. Thus we generalize the gauge approach to the symmetric teleparallel gravity so as to include both the nonmetricity and the curvature \cite{Pala2019}. Consequently we consider a Lagrangian which is quadratic in the nonmetricity and the full curvature, and also linear only in the full curvature scalar. We formulate our model in terms of the coordinate independent formalism, the so-called the exterior algebra in the orthonormal (Lorentzian) frame. We discuss the gauge group generated by the general linear coordinate transformation in the section \ref{sec:gen-lin-ccor-trans}.

With all this we add minimally the Dirac Lagrangian to our gravitational Lagrangian and vary the full Lagrangian in order to obtain field equations \cite{Kok2020}. From now on we restrict ourselves to two dimensions because  the investigation  of general 2D gravity models allows to tackle fundamental questions about quantum gravity by overcoming   challenges of higher dimensions. By studying the important physical properties of black holes in two dimensions especially with the string inspired models, valuable information can also be obtained for the systems in the quantum case. In the recent years, new insights have been gained for the exact quantization of the geometric part of such theories. They provide a systematic investigation for quantum field theoretical treatment, even when interacting with matter, without using a classical background geometry explicitly.  The review  on dilaton gravity in two dimensions \cite{grumiller2002} tries to assemble these results in a coherent manner, putting them at the same time into the perspective of the quite large literature on this subject. Again a new paper \cite{grumiller2021} focuses on a subclass of dilaton scale invariant models of the generalized dilaton gravity in 2D. Conversely spinor definition may have some subtle issues in two dimensions.

The explanation of special relativity is well served by the trick that we live in $1+1$ dimensional spacetime from many perspective, although rotations are impossible, boosts are colinear and some kinematic events  such as Thomas precession remain hidden. In the reference \cite{wheeler2000} the author finds that it has, in fact, many valuable lessons to impart, and that it speaks of deep things with engaging simplicity. But we know that mathematically spinors associated with 2D spacetime are defined as complex vectors carrying a representation of the covering group of the Lorentz group $SO(1,1)$. Additionally, the authors of the reference \cite{delhom2017} argue that they were able to  set a lower bound of the order of 1 TeV  on the scale at which  nonmetricity produces observable effects  in quantum fields without conflict with experiments by using available data for $e^+e^-$ scattering. Consequently we have enough motivation for investigating a toy model formulated in terms of the Dirac field, the nonmetricity and the curvature in a two dimensional spacetime.

In general a geometry is classified in accordance with the three tensors, the so-called the curvature, the torsion and the nonmetricity. If all three are zero, it is the Minkowskian geometry. If only the curvature is nonzero, the others are vanishing, it is Riemannian geometry. If only the nonmetricity is zero, the others are nonvanishing, it is called the Riemann-Cartan geometry. If only the torsion is nonzero, the others are vanishing, it is the Weitzenb\"{o}ck (or teleparallel) geometry. If only the nonmetricity is nonzero, the others are zero, it is called symmetric teleparallel geometry. If all three are nonzero, there is not a consensus on the nomenclature; the non-Riemannian geometry \cite{benn1981} or the metric affine geometry \cite{hehl1995}.

Finally we search some classes of stationary solutions to the field equations. We extensively make use of the computer algebra systems; REDUCE and MAPLE and  their packages EXCALC and ATLAS, respectively \cite{hearn2004}, \cite{schrufer2004}, \cite{Evans2009}.

\section{The mathematical preliminaries}
In spite of the fact that we will develop a theory in two dimensions in this work, all the content in this section is valid in $n$ dimensions. The spacetime, in general, is denoted by the triple $\{ M,g,\nabla\}$ where $M$ is the $n$-dimensional orientable and differentiable manifold, $g$ is the (0,2)-type symmetric and non-degenerate metric tensor, $\nabla$ is the connection representing the parallel transport of the tensors and spinors. Let $\{x^\alpha(p)\}, \ \alpha =\hat{0},\hat{1}, \cdots , \hat{n}-\hat{1}$, be the coordinate functions of the coordinate system at any point $p \in M$. This coordinate system forms the reference frame denoted by $ \{\frac{\partial}{\partial x^\alpha}(p)\}$ or shortly $\partial_\alpha (p) $, the so-called {\it coordinate (holonomic) frame}. This frame is a set of basis vectors at point $p$ for the tangent space $T_p(M)$. The union of all tangent spaces on $M$ is called the coordinate tangent bundle; $CT(M) = \bigcup_{p \in M}T_p(M)$. Similarly, the set of differentials of coordinate functions $\{ d x^\alpha (p) \}$ forms the {\it coordinate (holonomic) coframe} at the point $p$ for the cotangent space $T_p^*(M)$. Again the union of all $ T_p^*(M)$ establishes the coordinate cotangent bundle; $CT^*(M) = \bigcup_{p \in M}T_p^*(M)$. Duality between them is formulated by the relation
 \begin{equation}\label{duality1}
          d x^\alpha \left( \partial_\beta \right) = \delta^\alpha_\beta 
 \end{equation}
where $\delta^\alpha_\beta$ is the Kronecker symbol. In the coordinate frame the metric has the form
 \begin{equation}\label{metric1}
   g=g_{\alpha \beta}(x) dx^\alpha \otimes_S dx^\beta
 \end{equation}
where $\otimes_S$ denotes the symmetric tensorial product. We notice that the components of the metric tensor depend on the coordinates, $g(\partial_\alpha , \partial_\beta)=g_{\alpha \beta}(x)$.

On the other hand let $ \{X_a\},\ a=0,1,\cdots , n-1$, be the orthonormal set of basis vectors, the so-called {\it orthonormal (Lorentzian) frame}, such that the metric components are $g(X_a,X_b)=\eta_{ab}$ where $\eta_{ab}=\mbox{diag}(-1,1, \cdots , 1)$ is the Minkowski metric. The set $\{ X_a \}$ can be always obtained from the set $\{ \partial_\alpha \}$ or vice versa via the n-beine $h^\alpha{}_a$ or its inverse $h^a{}_\alpha $
  \begin{equation}\label{doublet1}
     X_a (x) = h^\alpha{}_a (x) \partial_\alpha \quad \leftrightarrow \quad \partial_\alpha = h^a{}_\alpha (x) X_a (x)
  \end{equation}
such that $h^\alpha{}_a(x) h^a{}_\beta(x) = \delta^\alpha_\beta$ and $h^a{}_\alpha (x) h^\alpha{}_b (x) = \delta^a_b$. Thus we can obtain the {\it orthonormal (Lorentzian) coframe} $\{ e^a \}$ through the duality relation
 \begin{eqnarray}\label{duality2}
          e^a(X_b) = \delta^a_b \ .
 \end{eqnarray}
This is other manifestation of the duality relation (\ref{duality1}). We can always pass from the orthonormal coframe to the coordinate coframe or vice versa by help of the n-beine defined in (\ref{doublet1}) as
  \begin{equation}\label{doublet2}
     dx^\alpha = h^\alpha{}_a(x) e^a(x) \quad \leftrightarrow \quad e^a(x) = h^a{}_\alpha(x) dx^\alpha \ .
  \end{equation}
While the set $\{ X_a(p) \}$ is the orthonormal basis of the tangent space $T_p(M)$, its dual $\{ e^a(p)\}$ is the orthonormal cobasis of the cotangent space $T_p^*(M)$ at point $p$ on $M$. Now as the union of all $ T_p(M)$ with $\{ X_a(p)\}$ establishes the orthonormal tangent bundle; $OT(M)$, the union of all $ T_p^*(M)$ with $\{ e^a(p)\}$ forms the orthonormal cotangent bundle; $OT^*(M)$. In the orthonormal frame the metric (\ref{metric1}) takes the form
 \begin{equation}\label{metric2}
   g = \eta_{ab} e^a(x) \otimes_S e^b(x) \ .
 \end{equation}
Here we pay special attention that the metric components $g(X_a,X_b)=\eta_{ab}$ are independent of the coordinates. 

In this work we use the language of the exterior algebra in which the coframe is called as the base 1-form or just 1-form. More explicitly $e^\alpha =dx^\alpha$ is the coordinate 1-form and $e^a$ is the orthonormal 1-form. Differential $d$ here is named as the exterior derivative in the exterior algebra converting the 0-form $x^\alpha$ to the 1-form $dx^\alpha$. Meantime the exterior derivative of the coframe is defined as the anholonomity 2-form. Therefore since $d(dx^\alpha)=0$ because of the Poincare lemma, $d^2=0$, $dx^\alpha$ is also known as the holonomic 1-form. However, the anholonomity of $e^a$ needs not to be zero, $de^a \neq 0$. Correspondingly, in the literature the coordinate indices are sometimes called as the holonomic indices and the orthonormal indices as the anholonomic indices. It should be noticed that in the coordinate frame $d(dx^\alpha)=0$, but $dg_{\alpha \beta} \neq 0$; in the orthonormal frame $de^a(x) \neq 0$, but $d\eta_{ab}=0$. Thus, apart from the coordinate and orthonormal basis it is always possible to work in a {\it mixed frame} in between them in which the exterior derivatives of both the metric components and the coframe are non-zero; $dg_{AB}(x) \neq 0$ and $de^A(x) \neq 0$.  Correspondingly the metric (\ref{metric1}) or (\ref{metric2}) may turn out to be
  \begin{equation}\label{metric3}
   g = g_{AB}(x) e^A(x) \otimes_S e^B(x)
 \end{equation}
where $A, B, \cdots = \bar{0}, \bar{1}, \cdots , \bar{n}-\bar{1}$ denotes the mixed indices. The orientation of the manifold is fixed by the Hodge map in terms of the orthonormal 1-forms; $*1=\frac{1}{n!} \epsilon_{a_1a_2 \cdots a_n} e^{a_1} \wedge e^{a_2} \wedge \cdots \wedge e^{a_n} = e^0 \wedge e^1 \wedge \cdots \wedge e^{n-1}$ where $\wedge$ denotes the exterior product. Here $\epsilon_{a_1 a_2 \cdots a_n}$ with the choice $\epsilon_{01\cdots (n-1)}=+1$ denotes the totally antisymmetric Levi-Civita tensor. From now on we make use of the abbreviation $e^{ab\cdots} \equiv e^a \wedge e^b \wedge \cdots$. Another important operation in the exterior algebra is the interior product, $\iota_{X_a} \equiv \iota_a$ or $\iota_{\partial_\alpha} \equiv \iota_\alpha$
 \begin{equation}\label{duality3}
   \iota_a e^b = \delta^b_a \quad \leftrightarrow \quad \iota_\alpha dx^\beta = \delta^\beta_\alpha
 \end{equation}
which are other manifestations of the duality relations (\ref{duality2}) and (\ref{duality1}), respectively. Here again $\iota_a$ and $\iota_\alpha$ are related via the n-beine, $\iota_\alpha = h^a{}_\alpha \iota_a$. Additionally the interior product of any 0-form is zero by definition. It satisfies a very useful identity together with the Hodge map; $*(\Phi \wedge e_a)=\iota_a *\Phi$ where $\Phi$ any $p$-form. The connection $\nabla$ is determined by the connection 1-forms ${\Lambda^a}_b$ by the relation $\nabla e^a = -e^b \otimes \Lambda^a{}_b$ where $\otimes$ denotes tensorial product. Under the transition between the coordinate and the orthonormal frames defined in (\ref{doublet1}) or equivalently (\ref{doublet2}), for any tensor-valued $p$-form to transform covariantly the connection 1-form must transform as
  \begin{equation}\label{doublet-con}
   \Lambda^a{}_b = h^a{}_\alpha \Lambda^\alpha{}_\beta h^\beta{}_b + h^a{}_\alpha dh^\alpha{}_b \quad \leftrightarrow \quad
   \Lambda^\alpha{}_\beta = h^\alpha{}_a \Lambda^a{}_b h^b{}_\beta + h^\alpha{}_a dh^a{}_\beta \ .
 \end{equation}
Thus we define the covariant exterior derivative of any $(p,q)$-type tensor-valued exterior form $\mathfrak{T}^{A_1A_2 \cdots A_p}_{\; \; \; \; B_1B_2 \cdots B_q }$ in a mixed frame as
 \begin{eqnarray}\label{covariant-derivative1}
   D \mathfrak{T}^{A_1A_2 \cdots A_p}_{\; \; \; \; B_1B_2 \cdots B_q } = d \mathfrak{T}^{A_1A_2 \cdots A_p}_{\; \; \; \; B_1B_2 \cdots B_q }
    + \Lambda^{A_1}{}_C \wedge \mathfrak{T}^{CA_2 \cdots A_p}_{\; \; \; \; B_1B_2 \cdots B_q } + \cdots + \Lambda^{A_p}{}_C \wedge \mathfrak{T}^{A_1A_2 \cdots C}_{\; \; \; \; B_1B_2 \cdots B_q } \nonumber \\
     - \Lambda^C{}_{B_1} \wedge \mathfrak{T}^{A_1A_2 \cdots A_p}_{\; \; \; \; CB_2 \cdots B_q } - \cdots
    - \Lambda^C{}_{B_q} \wedge \mathfrak{T}^{A_1A_2 \cdots A_p}_{\; \; \; \; B_1B_2 \cdots C} \ .
 \end{eqnarray}

The Cartan structure equations define the nonmetricity tensor 1-forms, the torsion tensor 2-forms and the curvature tensor 2-forms. They are written explicitly in a  mixed frame respectively as follows
 \begin{subequations}
  \begin{align}
      Q_{AB} &:= - \frac{1}{2} D g_{AB}
                      = \frac{1}{2} (-d g_{AB} + \Lambda_{AB}+\Lambda_{BA}) \ ,  \label{nonmet}\\
     T^A &:= D e^A = d e^A + {\Lambda^A}_B \wedge e^B \ ,  \label{tors}\\
     {R^A}_B &:= D {\Lambda^A}_B := d {\Lambda^A}_B + {\Lambda^A}_C \wedge {\Lambda^C}_B  \ . \label{curva}
  \end{align}
 \end{subequations}
They are not entirely independent because they satisfy the Bianchi identities
 \begin{subequations}
  \begin{align}
   D Q_{AB} &= \frac{1}{2} ( R_{AB} +R_{BA}) \ , \label{bianc:0} \\
       D T^A    &= {R^A}_B \wedge e^B \ ,  \label{bianc:1} \\
       D {R^A}_B &= 0 \ . \label{bianc:2}
   \end{align}
 \end{subequations}
The Cartan structure equations take the following forms in the coordinate frame
 \begin{subequations}
  \begin{align}
    Q_{\alpha \beta} &= \frac{1}{2} (-d g_{\alpha \beta} + \Lambda_{\alpha \beta}+\Lambda_{\beta \alpha}) \ ,  \label{nonmet1}\\
    T^\alpha  &= {\Lambda^\alpha}_\beta \wedge dx^\beta \ , \label{tors1}\\
    {R^\alpha}_\beta &= d {\Lambda^\alpha}_\beta + {\Lambda^\alpha}_\gamma \wedge {\Lambda^\gamma}_\beta \ , \label{curva1}
   \end{align}
 \end{subequations}
and those in the orthonormal frame
   \begin{subequations}
 \begin{align}
  Q_{ab} &= \frac{1}{2} (\Lambda_{ab}+\Lambda_{ba}) \ , \label{nonmet2}\\
    T^a &= d e^a + {\Lambda^a}_b \wedge e^b \ ,  \label{tors2}\\
    {R^a}_b &= d {\Lambda^a}_b + {\Lambda^a}_c \wedge {\Lambda^c}_b \ . \label{curva2}
   \end{align}
 \end{subequations}
Their geometric meanings become obvious when we think of parallel transport of a vector along a closed loop on $M$. After one complete revolution of the vector if its final length is not the same as the initial length, then there is nonmetricity; if there is an angle between the final and the initial vectors, then there is curvature; and if there is a shift between the final and the initial vectors in the tangent bundle (or if the contour image of the loop in the tangent bundle broken), then there is torsion.

\subsection{Decomposition of the full connection } \label{sec:connec-decomp}

In a {\it mixed frame} the full connection 1-form can be decomposed uniquely as follows \cite{hehl1995},\cite{benn1982},\cite{tucker1995}
 \ba
     {\Lambda^A}_B =  \underbrace{(g^{AC}d g_{CB} + {p^A}_B)/2 + {\omega^A}_B}_{Metric}
     + \underbrace{{K^A}_B}_{Torsion} + \underbrace{{q^A}_B  + {Q^A}_B}_{Nonmetricity}   \label{connect:dec}
 \ea
where ${\omega^A}_B$ is the Levi-Civita connection 1-form
 \ba
     {\omega^A}_B \wedge e^B = -d e^A \ ,   \label{LevCiv}
 \ea
$K^A{}_B$ is the contortion tensor 1-form,
 \ba
   {K^A}_B  \wedge e^B = T^A  \ , \label{contort}
 \ea
$p_{AB}$ and $q_{AB}$ are defined as
 \ba
  & & p_{AB} = -( \imath_A d g_{BC} ) e^C + ( \imath_B d g_{AC}) e^C \ , \label{p:ab} \\
   & & q_{AB} = -( \imath_A  Q_{BC} ) e^C + ( \imath_B Q_{AC}) e^C  \ . \label{q:ab}
 \ea
This decomposition is self-consistent. To see that it is enough to multiply (\ref{connect:dec}) from right by $\wedge e^B$ and to use the definitions above. While moving indices vertically in front of both $d$ and $D$, a special attention is needed because $d g_{AB} \neq 0$ and $D g_{AB} \neq 0$. The symmetric part of the full connection comes from (\ref{nonmet})
 \ba
  \Lambda_{(AB)} = Q_{AB } + \frac{1}{2} d g_{AB } \label{connect:sym}
 \ea
and the remainder is the anti-symmetric part
 \ba
  \Lambda_{[AB]} = \frac{1}{2} p_{AB} + \omega_{AB} + K_{AB} + q_{AB} \ .  \label{connect:ansym}
 \ea
If only $Q_{AB}=0$, the connection is said to be metric compatible. If both $Q_{AB}=0$ and $T^A =0$, the connection becomes the Levi-Civita. In the coordinate (holonomic) frame, with the abbreviation $\iota_\alpha d \equiv \partial_\alpha$, the decomposition (\ref{connect:dec}) reduces to
  \begin{equation}\label{decomp-con-coord}
{\Lambda^\alpha}_\beta =  \underbrace{\frac{1}{2}g^{\alpha \sigma}(\partial_\gamma g_{\sigma \beta} + \partial_\beta g_{\sigma \gamma} - \partial_\sigma g_{\beta \gamma})dx^\gamma}_{Metric}
     + \underbrace{{K^\alpha}_\beta}_{Torsion} + \underbrace{{q^\alpha}_\beta  + {Q^\alpha}_\beta}_{Nonmetricity}
 \end{equation}
where the first group on the right hand side is the Christoffel symbols. In the orthonormal (Lorentzian) frame it takes the form
 \ba
 \Lambda_{ab} = \omega_{ab} + K_{ab} + q_{ab}
          +  Q_{ab} \ . \label{connect:on}
 \ea
Here the Levi-Civita connection 1-form is solved in terms of orthonormal coframe
 \begin{equation}\label{Levi-Civita:on}
   \omega_{ab} = \frac{1}{2} \left[ -\iota_a de_a + \iota_b de_a + (\iota_a \iota_b de_c) e^c \right]
 \end{equation}
and the contortion 1-form in terms of the torsion
   \begin{equation}\label{contorsion:on}
   K_{ab} = \frac{1}{2} \left[ \iota_a T_b - \iota_b T_a - (\iota_a \iota_b T_c) e^c \right] \ .
 \end{equation}
Besides the quantity $q_{ab}$ is defined in terms of the nonmetricity
 \begin{equation}\label{qab-orthnorm}
   q_{ab} = -( \imath_a  Q_{bc} ) e^c + ( \imath_b Q_{ac}) e^c \ .
 \end{equation}
In the literature there are works preferring the coordinate frame \cite{hehl1995}, the orthonormal frame \cite{dereli1994} and a mixed frame \cite{adak2006}. In calculations the following identities are useful:
  \begin{align}
    D*e_{a_1} &= - Q \wedge *e_{a_1} + * e_{a_1a_2} \wedge T^{a_2} \ , \nonumber \\
    D*e_{a_1a_2} &= -  Q \wedge *e_{a_1a_2} + * e_{a_1a_2a_3} \wedge T^{a_3} \ , \nonumber \\
   \vdots & \label{ozedeslik D hodge kocerceve} \\
 D*e_{a_1 a_2 \cdots a_{n-1}}  &= -  Q \wedge *e_{a_1 a_2 \cdots a_{n-1}} + e_{a_1 a_2 \cdots a_{n-1}a_n} \wedge T^{a_n} \ , \nonumber \\
   D*e_{a_1 a_2 \cdots a_n}  &= -  Q \wedge *e_{a_1 a_2 \cdots a_n} \ , \nonumber \\
    D\eta_{ab} &= -2Q_{ab} \, , \quad D\eta^{ab} = +2Q_{ab} \, , \quad D\delta^a_b =0 \nonumber
   \end{align}
where $ Q := Q^a{}_a = \eta^{ab} Q_{ab} = \Lambda^a{}_a$ is the trace 1-form of nonmetricity.

\subsection{General linear coordinate transformation} \label{sec:gen-lin-ccor-trans}

In this subsection we consider the general linear coordinate transformation, $x^\mu \rightarrow x^{\mu'}$ in the form
 \begin{equation} \label{gen-lin-coord-trans}
   x^{\mu'} = \Gamma^{\mu'}{}_\mu x^\mu + \xi^{\mu'}
 \end{equation}
where $ \Gamma^{\mu'}{}_\mu = \Gamma^{\mu'}{}_\mu (x)$ represents the rotational part and $\xi^{\mu'}=\xi^{\mu'}(x)$ translational part. Now we take the exterior derivative of the both sides which is coordinate independent by definition.
  \begin{equation}
   dx^{\mu'} = \Omega^{\mu'}{}_\mu dx^\mu 
 \end{equation}
where $\Omega^{\mu'}{}_\mu(x) := [\partial_\mu \Gamma^{\mu'}{}_\nu (x)]x^\nu + \Gamma^{\mu'}{}_\mu(x) + \partial_\mu\xi^{\mu'}(x)$. It is worthwhile to remark that the transformation elements $\Omega^{\mu'}{}_\mu(x)$ form the general linear group, $Gl(n,\mathbb{R})$. We pass to the orthonormal coframe in order to see the effects of the general linear coordinate transformations on it by help of the n-beine
 \begin{equation}
  dx^{\mu'} = h^{\mu'}{}_{a'} e^{a'} \quad \mbox{and} \quad dx^{\mu} = h^{\mu}{}_{a} e^{a}
 \end{equation}
where $ h^{\mu'}{}_{a'} =h^{\mu'}{}_{a'} (x'(x))$ and $h^{\mu}{}_{a}=h^{\mu}{}_{a}(x)$. After substituting these into the equation and then by using the relation $h^{b'}{}_{\mu'}h^{\mu'}{}_{a'} = \delta^{b'}_{a'}$ we obtain
  \begin{equation}\label{oncoframe-transf}
   e^{b'} = L^{b'}{}_b e^b
  \end{equation}
where $L^{b'}{}_b (x'(x)) := h^{b'}{}_{\mu'} \Omega^{\mu'}{}_\mu  h^{\mu}{}_b$. It has to be noticed that the transformation of the orthonormal coframe generated by the general linear coordinate transformation defined by (\ref{gen-lin-coord-trans}) does not necessarily have a translational piece. Finally we want to look at the transformation rule for the metric in the orthonormal frame
 \begin{equation}
   g= \eta_{ab}e^a \otimes e^b = \eta_{a'b'} e^{a'} \otimes e^{b'}
  \end{equation}
where $\eta_{ab}=\eta_{a'b'}= \mbox{diag}(-1,+1, \cdots , +1)$ are the Minkowski metric. Here $e^{a'} = L^{a'}{}_a e^a$ induces $ \eta_{a' b'}=L^a{}_{a'} \eta_{ab} L^b{}_{b'}$. In the matrix notation it is read as
 \begin{equation}
   e' = L^{-1} e \quad \rightarrow \quad \eta' = L^T \eta L
 \end{equation}
where $^T$ denotes the transpose matrix. This shows that the transformation elements obtained from the general linear coordinate transformation generate the Lorentz group $SO(1,n-1)$ in the orthonormal bundle. That is why one calls orthonormal 1-forms as the Lorentzian coframe as well.  Thus in the orthonormal bundle while the orthonormal coframe transforms by the rule (\ref{oncoframe-transf}), the full connection must transform according to the rule
 \begin{equation}\label{onconnection-transf}
   \Lambda^{a'}{}_{b'} = L^{a'}{}_a \Lambda^a{}_b L^b{}_{b'} + L^{a'}{}_a dL^a{}_{b'}
 \end{equation}
such that the nonmetricity tensor 1-form, the torsion tensor 2-form and the curvature tensor 2-form can transform covariantly
 \begin{subequations}
  \begin{align}
    Q_{a' b'} &= L^a{}_{a'} Q_{ab} L^b{}_{b'} \ , \label{onQ-transf}\\
    T^{a'} &= L^{a'}{}_a T^a \ , \label{onT-transf}\\
    R^{a'}{}_{b'} &= L^{a'}{}_a R^a{}_b L^b{}_{b'} \ . \label{onR-transf}
  \end{align}
 \end{subequations}
We conclude this subsection by stating that the gauge group of the orthonormal bundle is the Lorentz group independent of whether the nonmetricity is or not in the spacetime. This conclusion agrees with that of the reference \cite{benn1982}.


\section{The Lagrange formulation of a theory}
In the Lagrangian formalism of a gravity theory, firstly an action is written containing a Lagrangian
 \begin{equation}\label{action}
   I=\int_M L
 \end{equation}
where $L$ is called the Lagrangian $n$-form. Extremum of this action yields the field equations. In other words, the variation of the action is set to zero in order to obtain the field equations; $\delta I =0$ meaning $\delta L=0$. From now on we use the Lagrangian and the Lagrangian $n$-form interchangeably. In general, the Lagrangian contains geometric and matter parts. In the orthonormal frame the independent variables of a Lagrangian are only the orthonormal coframe, $e^a$, the full connection 1-form, ${\Lambda^a}_b$, and a matter field $\psi$ but not the metric components because $\delta \eta_{ab}=0$. In this case the variation of the Lagrangian, $L=L[e^a,{\Lambda^a}_b, \psi]$, is written as
 \begin{equation}
  \delta L = \delta e^a \wedge \frac{\partial L}{\partial e^a}
  + \delta \Lambda_{ab} \wedge \frac{\partial L}{\partial \Lambda_{ab}}
  + \delta \psi \wedge \frac{\partial L}{\partial \psi} \ .
 \end{equation}
In the literature it is shown as $\frac{\partial L}{\partial e^a} := \tau_a$ and called the energy momentum $(n-1)$-form and similarly $\frac{\partial L}{\partial \Lambda_{ab}}:= \Sigma^{ab}$ the angular momentum $(n-1)$-form. In MAG formulation $\Sigma^{ab}$ is called the hypermomentum $(n-1)$-form \cite{hehl1995}.

\subsection{Form of Lagrangian}

Since the gauge theory works very well in the microscopic world both theoretically and experimentally, we allow it to guide us for searching a modified theory of gravity. Therefore we adhere the arguments in the reference \cite{adak2018} and summarize them in the Table \ref{tab:analogy} in which $A$ is called the potential 1-form and $F$ the Maxwell 2-form. For the meanings of other symbols, please consult for the next subsection.  Accordingly any gravitational Lagrangian should contain some quadratic nonmetricity terms representing gravitational field and quadratic curvature terms as gauge fields.  
 \begin{table}[ht]
 \caption{Analogy between the theory of gravity and the Maxwell-Dirac theory. Because of the discussions in the subsection \ref{sec:gen-lin-ccor-trans}, we would like to emphasize that the gauge group $Gl(n,\mathbb{R})$ reduces to the Lorentz group in the orthonormal frame.}
 \centering
 \renewcommand{\arraystretch}{1.7} 
\begin{tabular}{ |p{0.25\textwidth}|p{0.33\textwidth}|p{0.33\textwidth}| }
\hline
Description & Maxwell-Dirac & Gravity \\
\hline
\hline
Physical field & $\psi (x)$ & $g_{\alpha \beta}(x)  $ \\
\hline
Globally gauge invariant Lagrangian & $\frac{i}{2} ( \overline{\psi} *\gamma \wedge d\psi + d\overline{\psi} \wedge *\gamma \psi ) + i m \overline{\psi} \psi *1 $ & $\frac{\kappa}{8} dg_{\alpha \beta} \wedge *dg^{\alpha \beta} + \frac{M}{4} g_{\alpha\beta} g^{\alpha \beta}*1$\\
\hline
Global gauge transformation & $\psi \rightarrow e^{i\theta} \psi$ where $\theta$ is any real number & $g \rightarrow \Omega^T g \Omega$ where $\Omega$ contains real numbers \\
\hline
Locally gauge invariant Lagrangian & $\frac{i}{2}( \overline{\psi} *\gamma \wedge \mathcal{D} \psi + \mathcal{D}\overline{\psi} \wedge *\gamma \psi) +i m \overline{\psi} \psi *1 + F \wedge *F $ \newline where \newline $\mathcal{D} \psi := (d-iA)\psi$, \newline $F:=dA$ & $\frac{\kappa}{8} Dg_{\alpha \beta} \wedge *Dg^{\alpha \beta} + M*1 + {R^\alpha}_\beta \wedge *{R^\beta}_\alpha$ \newline where \newline $Dg_{\alpha \beta} := \frac{1}{2} (-d g_{\alpha \beta} + \Lambda_{\alpha \beta}+\Lambda_{\beta \alpha})$, ${R^\alpha}_\beta  := d {\Lambda^\alpha}_\beta
                  + {\Lambda^\alpha}_\gamma \wedge {\Lambda^\gamma}_\beta $ \\
\hline
Local gauge transformations & $\psi \rightarrow e^{i\theta(x)} \psi$, \newline  $A \rightarrow A+d\theta$ & $g \rightarrow \Omega^T(x) g \Omega(x)$, \newline $\Lambda \rightarrow \Omega^{-1} \Lambda \Omega + \Omega^{-1} d \Omega$  \\
\hline
Gauge group & $U(1)$ & $Gl(n,\mathbb{R})$\\
\hline
Gauge field (potential) & $A$ & $\Lambda^\alpha{}_\beta $\\
\hline
Bianchi identities & $\mathcal{D}^2\psi =-iF\psi$,  \newline $ dF=0$ & $D^2g_{\alpha \beta} = -( R_{\alpha \beta} + R_{\beta \alpha})$, \newline $D {R^\alpha}_\beta=0$\\
\hline
\end{tabular}
\label{tab:analogy}
 \end{table}

\subsection{Variational field equations}

Up to this point all developments and arguments are valid in any dimensions. But from now on we restrict ourselves to only two dimensions because the investigation  of general 2D gravity models allows to tackle fundamental questions about quantum gravity by overcoming challenges and technical complications  of higher dimensions. In the table 1 we write consciously all tensor quantities in the coordinate frame with the holonomic indices to make especially $dg_{\alpha \beta}$ clearly visible. On the other hand, in the language of the exterior algebra we always write a Lagrangian in $OT(M)$ because it is independent of choice of coordinate chart. Thus, any alternative theory is definitely invariant under a general coordinate transformation as we discussed in the subsection \ref{sec:gen-lin-ccor-trans}. However, whenever we try to find some exact solutions, we choose a coordinate system and then pass to $CT(M)$ via n-beine. Along with these since we also couple a Dirac field to gravity we must go to the spinor bundle and we can pass to there through $OT(M)$. Consequently we consider the following Lagrangian 2-form as our gravity model
 \begin{eqnarray} \label{total lagrangian}
   L= \kappa R^{a}{_{b}} \wedge * e_a{}^b  +\mu R^a{}_b \wedge * R^b{}_a +\nu Q_{ab} \wedge * Q^{ab} + \lambda_{a} T^{a} + \Lambda *1 \nonumber \\
   + \frac{i}{2} \left[ \overline{\psi} *\gamma \wedge (D \psi) + (D \overline{\psi}) \wedge * \gamma \psi \right] + \frac{imc}{\hbar} \overline{\psi} \psi *1
 \end{eqnarray}
where $\kappa, \mu, \nu$ are coupling constants, $\Lambda$ is the cosmological constant, $\lambda_a$ is the Lagrange multiplier 0-form constraining torsion to zero, $m$ is the mass of the Dirac particle, $c$ is the speed of light in vacuum, $\hbar$ is the Planck constant and $i$ is the complex unit $i^2=-1$. The first line of the Lagrangian represents the geometric (gravitational) part and the second line is the Dirac Lagrangian 2-form. The first term of the first line is the Einstein-Hilbert term in a non-Riemannian form. While we are discussing Yang-Mills-type prescription (quadratic in both curvature and nonmetricity), we included it because when we want to obtain the Einstein limit by making nonmetricity zero, we need the linear term in curvature in the Lagrangian in order to obtain the Einstein equations. A similar Lagrangian in which torsion and nonmetricity exchanged has been studied in \cite{tseytlin1982} where the author remarks that the standard Yang-Mills formulation of Poincare gravity does not give the correct results. The covariant exterior derivative of the Dirac spinor is given by
 \begin{eqnarray} \label{covariant exterior derivative of psi}
    D \psi= d \psi + \frac{1}{2} \Lambda^{[ab]} \sigma_{ab} \psi + b Q \psi 
 \end{eqnarray}
where $b$ is a constant (complex or real). According to the authors of the paper \cite{janssen2018} if $b$ is real-valued, it is related with local rescalings of the spinor while if it has purely imaginary values, it corresponds to local $U(1)$ transformations of the spinor. Here since the Dirac matrices, $\gamma_0$ and $\gamma_1$,  satisfy the anti-commutation relation $ \{ \gamma_a , \gamma_b \} = 2 \eta_{ab} I$ where $I$ is $2 \times 2$ unitary matrix, they generate $ \mathcal{C}\ell_{1,1}$ Clifford algebra. We choose the Dirac matrices as
\begin{eqnarray} \label{dirac matrices}
\gamma_{0}= \begin{pmatrix}
0 & -1 \\ 
1 & 0
\end{pmatrix}  & \text{and} & \gamma_{1}= \begin{pmatrix}
1 & 0 \\ 
0 & -1
\end{pmatrix} \ .
\end{eqnarray}
In this representation the Dirac spinor becomes a two-component column matrix. Dirac adjoint of $\psi$ field is defined as $\overline{\psi}:= \psi^\dagger \gamma_0$. Accordingly,
\begin{eqnarray} \label{covariant exterior derivative of psi line}
D \overline{\psi} = d \overline{\psi} - \frac{1}{2} \overline{\psi} \sigma_{ab} \Lambda^{[ab]} + b^{\star} \, Q \overline{\psi}
\end{eqnarray}
and $\gamma = \gamma_a e^a$ is called the $\mathcal{C}\ell_{1,1}$-valued 1-form where $^\star$ denotes the complex conjugation, $i^\star =-i$. Besides $\sigma_{ab} := \frac{1}{4}[\gamma_a , \gamma_b] $ is the generator of the Lorentz group, $SO(1,1)$. Thus, spinors associated with $M_2$ are defined as complex vectors carrying a representation of
the double covering group of $SO(1,1)$.

Now we vary the Lagrangian with respect to $\lambda_a$, $e^a$, $\Lambda^a{}_b$ and $\overline{\psi}$, respectively and obtain the field equations below
 \begin{subequations} \label{field equations}
 \begin{align}
     T^a &= 0 \ , \label{zero torsion} \\
     D\lambda_a + \tau_a[R] + \tau_a[Q] + \tau_a[\psi] + \Lambda *e_a &=0 \ , \label{coframe equation} \\
     \lambda_a e^b + \Sigma_a{}^b[R] + \Sigma_a{}^b[Q] + \Sigma_a{}^b[\psi] &=0 \ , \label{connection equation} \\
       *\gamma \wedge \left\{ D +\frac{1}{2} \left[ (\iota_b Q^{ab}) e_a  -(1+b+b^\star) Q\right] \right\} \psi + \frac{mc}{\hbar} \psi *1  &=0 \label{dirac equation}
 \end{align}
 \end{subequations}
where we defined the energy-momentum 1-forms, $\tau_a[\cdot]$, 
 \begin{subequations}
 \begin{align} 
 \tau_a[R] &:= - \mu  (\iota_{a} R^{c}{_b}) \wedge * R^{b}{_c} \ ,  \\
 \tau_a[Q] &:= - \nu  [(\iota_{a} Q^{bc}) \wedge * Q_{bc} + Q^{bc} \wedge (\iota_{a} * Q_{bc})] \ , \\
 \tau_a[\psi] &:= \frac{i}{2} \left[ \overline{\psi} *(\gamma \wedge e_{a}) \wedge (D \psi) - (D \overline{\psi}) \wedge *(\gamma \wedge e_{a}) \psi \right] + \frac{imc}{\hbar} \overline{\psi} \psi *e_{a} \ , \label{psienergymomentum}
 \end{align}
 \end{subequations}
and the angular momentum (or hypermomentum) 1-forms, $\Sigma_a{}^b[\cdot]$, 
 \begin{subequations}
 \begin{align}
 \Sigma_a{}^b[R] &:=  2\mu D * R^{b}{_a} \ , \\
 \Sigma_a{}^b[Q] &:= \kappa [2Q^{bc} \wedge * e_{ac} - Q \wedge * {e_a}^b] +  2 \nu  * {Q_a}^b \ , \\
 \Sigma_a{}^b[\psi] &:= \frac{i(b^{\star} - b)}{2} \delta_a^b \overline{\psi} *\gamma \psi \label{psiangularmomentum}
 \end{align}
 \end{subequations}
coming from the curvature, the nonmetricity, the spinor, respectively. It is worthy to notice that as $b$ is a real constant, angular momentum of the spinor vanishes. Here we also used the abbreviation $D*R^b{}_a := d *R^b{}_a + \Lambda^b{}_c \wedge *R^c{}_a - \Lambda^c{}_a \wedge *R^b{}_c$. The connection equation (\ref{connection equation}) is an algebraic equation for the lagrange multiplier $\lambda_a$ and in the language of exterior algebra we can calculate $\lambda_a$ uniquely by applying $\iota_b$ to it
 \begin{eqnarray} \label{lambda-a}
  \lambda_a= -\frac{1}{2}\iota_b \left(\Sigma_a{}^b[R] + \Sigma_a{}^b[Q] + \Sigma_a{}^b[\psi]\right) .
 \end{eqnarray}
Now the result (\ref{lambda-a}) has to be inserted in the equation (\ref{coframe equation}). Thus in order to find a class of solution to our theory we must make ansatz for $e^a$, $Q_{ab}$ and $\psi$ and they must satisfy the equations (\ref{zero torsion}), (\ref{coframe equation}) and (\ref{dirac equation}). We also insert (\ref{lambda-a}) back to (\ref{connection equation}) and obtain a constraint between the $\Sigma$'s, $2\Sigma_a{}^b - (\iota_c \Sigma_a{}^c)e^b=0$,
where $\Sigma_a{}^b \equiv \Sigma_a{}^b[R] + \Sigma_a{}^b[Q] + \Sigma_a{}^b[\psi] $.


\section{Some classes of stationary solutions}

In general there are 16 unknowns (4 from $e^a$ plus 8 from $\Lambda^{ab}$ plus 2 from $\lambda_a$ plus 2 from $\psi$) in our theory. Consistently we obtained 16 equations (2 from (\ref{zero torsion}) plus 4 from (\ref{coframe equation}) plus 8 from (\ref{connection equation}) plus 2 from (\ref{dirac equation})). Since we were not successful to find an exact stationary solution to our field equations, we decided to follow the strategy. Correspondingly we firstly omit the contribution of $\psi$ to gravity, i.e., set $\tau_a[\psi]=0$ and $\Sigma^a{}_b[\psi]=0$ in the equations  (\ref{coframe equation}) and (\ref{connection equation}) and then obtain some classes of vacuum solutions to them. Finally we solve the Dirac equation for a spinning particle in that gravitational field. Therefore we assume a gravitation that is described by the following configuration in the $x^\mu = (t,x)$ coordinate chart
 \begin{eqnarray}
 & &  e^0 = c f(x) dt \, , \quad e^1 = g(x) dx \ , \label{ansatz ea}\\
 & &  Q_{00} = h_1(x) dx \ , \quad Q_{11} = h_2(x) dx \ , \quad Q_{01} = Q_{10} = 0 \label{ansatz Qab}
 \end{eqnarray}
where $f(x), g(x), h_1(x), h_2(x)$ are real unknown functions. From the equation (\ref{ansatz Qab}) we deduce $Q = [h_2(x) - h_1(x)] dx$. 

We calculate the Levi-Civita connection 1-forms by inserting (\ref{ansatz ea}) to (\ref{Levi-Civita:on})
 \begin{eqnarray}
  \omega_{01} = -\omega_{10} = -\frac{f'}{fg}e^0 \, , \quad \omega_{00}=\omega_{11}=0 \label{omega01}
 \end{eqnarray}
and $q_{ab}$ putting (\ref{ansatz Qab}) to (\ref{qab-orthnorm})
  \begin{eqnarray}
  q_{01} = -q_{10} = \frac{h_1}{g}e^0 \, , \quad q_{00}=q_{11}=0 \label{q01}
 \end{eqnarray}
where the prime denotes the derivative with respect to $x$. By combining (\ref{ansatz Qab}), (\ref{omega01}) and (\ref{q01}) in (\ref{connect:on}) together with $K_{ab}=0$ because of $T^a=0$ we could write down the full connection 1-forms
  \begin{eqnarray}
  \Lambda_{01} = -\Lambda_{10}= -\frac{f' - f h_1}{fg}e^0 \, , \qquad \Lambda_{00} = \frac{h_1}{g}e^1 , \quad \Lambda_{11} = \frac{h_2}{g}e^1 \ . \label{Lambda01}
 \end{eqnarray}
Now we have checked that the zero torsion constraint, (\ref{zero torsion}), is satisfied. Then the full curvature 2-forms could be calculated by using the definition (\ref{curva2}). 

Thus for obtaining a class of exact solution we assume the following relations among the unknown functions
\begin{equation}
    h_1(x) = a_1 f'/f , \quad h_2(x) = a_2 h_1 , \quad fg = 1
\end{equation}
where $a_1$ and $a_2$ are arbitrary constants. Accordingly we obtained the following
\begin{equation}
    f = C_1 + C_0 x, \quad g = \frac{1}{C_1 + C_0 x}, \quad h_1 = \frac{a_1 C_0}{C_1 + C_0 x}, \quad h_2 = \frac{a_1 a_2 C_0}{C_1 + C_0 x}  \label{ref:sol_geometric_part}
\end{equation}
where $C_1$ is an arbitrary constant of integration. The crucial constant $C_0$ is given by the following expression, in terms of the coupling constants $\kappa$, $\mu$ and $\nu$, and the constants $a_1$ and $a_2$ that determine the nonmetricity: 
\begin{equation}
    C_0^2 = \frac{\kappa(2a_1 - 1)(a_2 + 1) - 2\nu(2a_1a_2^2 + 1)}{2\mu(a_2 + 1)\left[2a_1^3(a_2 - 1) - 3a_1^2(a_2 - 3) + a_1(a_2 - 10) + 3 \right]} \ . \label{C0 constant}
\end{equation}
It is worthwhile to remark some points. 
\begin{enumerate}[i.]
    \item The case of $C_0=0$ yields flat spacetime. Correspondingly we might conclude that the constant $C_0$ behaves delicately like an implicit source of the gravity.
    \item By rewriting explicitly the nonmetricity and the full curvature
\begin{subequations}
\begin{align}
     Q_{01} &= Q_{10} =0 \, , \quad Q_{00} = C_0 a_1 e^1 \, , \quad Q_{11} = C_0 a_1a_2 e^1 \ ,  \\
     R^0{}_0 &= R^1{}_1 = 0 \, , \quad R^0{}_1 = -C_0^2 \left[a_1(a_2 + 1) - 1 \right](a_1 - 1)e^{01} \, , \nonumber \\  R^1{}_0 &= C_0^2 \left[a_1(a_2 + 1) + 1 \right](a_1 - 1)e^{01} 
\end{align}
\end{subequations}
    we can gradually trace gravitational contributions resulting from the nonmetricity and the Riemannian curvature. That is, $a_2=0$ kills only $Q_{11}$, then $a_1=0$ kills the remainder of $Q_{ab}$ and at the end $\kappa + 2 \nu=0$ kills the Riemannian contribution to gravity.
    \item This solution is not valid for $a_2 = -1$ or $\mu = 0$ because $C_0$ becomes infinity.
    \item There are three singular points, namely $x=-C_1/C_0$ and $x=\pm \infty$. These singularities may be replaced with $x=-C_0/C_1$ and $x=0$ by the coordinate transformation $x \rightarrow 1/x$. For gaining an insight for the character of these singularities we calculate the following scalars 
    \begin{subequations}\label{singularity analysis}
     \begin{align}
         *(Q_{ab} \wedge *Q^{ab}) &= -C_0^2 a_1^2 (1 + a_2^2) \ , \\
         *(R^a{}_b \wedge *R^b{}_a) &= -2C_0^4 [(a_1^2 a_2 + a_1^2 - a_1 a_2 - 2a_1)^2 - 1] \ .
     \end{align}
    \end{subequations}
    Accordingly although not definite, the singular points look like to be coordinate singularities because both scalars are finite at the singular points. In the Riemannian spacetimes, test particles trace the geodesics of the geometry and there are singularity theorems that show the existence of incomplete trajectories of those test particles moving in the Einsteinian spcatimes. All these discussions are ended with existence of black holes \cite{Hawking1970}, \cite{Hawking1973}, \cite{Wald1984}. However, instead of these steps, in practice we look at the singularities of metric functions and the Riemannian curvature scalar, and then we deduce if there is a black hole or not. On the other hand, in this work we are in a non-Riemannian spacetime containing nonmetricity and we do not have theorems about the existence of incomplete trajectories followed by test particles. Even we are not sure on which curve a test particle follow in a non-Riemannian spacetime \cite{adak2011}. Therefore we can not conclude that our solution describes a black hole or not. 
    \item As $a_1 = 0$, nonmetricity vanishes and the theory reduces a Riemannian theory. By setting $fg = 1$ we obtain two more classes of solutions
\begin{subequations}
\begin{align}
    f &= \left[D_2 + D_1 x \pm\left( -\frac{\Lambda}{2\mu} \right)^{1/2} x^2 \right]^{1/2} \ , \\
    f &= D_1 + \left( -\frac{\Lambda}{2\mu} \right)^{1/4} x
\end{align}
\end{subequations}
where $D_1$ and $D_2$ are integral constants. 
    \item As $a_1 = 1$, the full curvature vanishes (but the Riemannian curvature is nonzero) in which case the theory turns out to be a symmetric teleparallel gravity model. We observe the same results as those in \cite{adak2008}. Additionally we obtain two more classes of solutions
 \begin{itemize}
     \item With the adjustments of $\kappa = 2\nu$, $a_2 \neq 0$ and $g=1/f$ 
     \begin{subequations}
\begin{align}
    h_1 = f'/f, \qquad h_2 = \frac{1}{f}\left(\frac{\Lambda}{\nu} - f'\,^2 \right)^{1/2} \ , \\
    f = D_0 \pm\sqrt{\frac{4\Lambda}{5\nu}}x \qquad \text{or} \qquad f = \left(D_1 + D_0 x + \frac{\Lambda}{\nu}x^2 \right)^{1/2}
\end{align}
\end{subequations}
where $D_0$ and $D_1$ are arbitrary integral constants. 
 \item With the adjustments of $\kappa \neq 2\nu$, $ a_2=0$ and $g=1$ 
 \begin{align}
       h_1 = f'/f \ , \qquad h_2=0 \ , \qquad
      f = (D_1 + D_0x)^{\displaystyle\frac{\kappa+2\nu}{\kappa-2\nu}}
 \end{align}
again here $D_0$ and $D_1$ are arbitrary constants.
 \end{itemize}
\end{enumerate}

\subsection{Hamiltonian of a spinning particle in the gravitational field}

At this point we want to deduce the Dirac Hamiltonian for a spinning particle in the gravitational field described by the solution (\ref{ref:sol_geometric_part}). Thus we substitute our result (\ref{ref:sol_geometric_part}) into the Dirac equation (\ref{dirac equation}) and reach the explicit form of the Dirac equation
\begin{equation}\label{ref:Dirac_eq_like_Sch}
    i\hbar\frac{\partial\Psi}{\partial t} = \left( i\gamma_0 mc^2 f + i \gamma_0\gamma_1 \hbar cNf - \gamma_0\gamma_1 cf^2 i\hbar\frac{\partial}{\partial x} \right)\Psi
\end{equation}
where 
\begin{equation}
    N := \frac{C_0}{2}\left[ a_1(a_2 - 1)(b - b^*) + 1 \right] \ . \label{N definition}
\end{equation}
Here we define canonical momenta and assume that it is hermitian
\begin{equation}\label{ref:momentum_operator_general}
    \hat{p} := -i\hbar\left( \frac{\partial}{\partial x} - \frac{N}{f} \right) \ .
\end{equation}
Consequently the Dirac equation turns out to be
\begin{equation}
    i\hbar\frac{\partial\Psi}{\partial t} = \left( i\gamma_0 mc^2 f + \gamma_0\gamma_1 cf^2 \hat{p} \right)\Psi.
\end{equation}
By comparing this with the Schrödinger equation $i\hbar \partial \psi / \partial t= \hat{H} \psi$ we deduce the Hamiltonian matrix
\begin{align}
    \hat{H} &= f\left( i\gamma_0 mc^2 + \gamma_0\gamma_1 cf\hat{p} \right) = \begin{pmatrix} 0 & f\left( - imc^2 + cf\hat{p} \right) \\ f\left( imc^2 + cf\hat{p} \right) & 0 
     \end{pmatrix}  .
\end{align}
We compute the eigenvalues of this $2\times 2$ hermitian matrix as $\pm f\left(m^2c^4 + f^2 p^2 c^2 \right)^{1/2}$ 
where $p$ is the eigenvalue of $\hat{p}$. We remark that as $C_0$ goes to zero, gravitational field vanishes that means $f=1$. In that case our eigenvalues reduces to the well known values; $\pm \left(m^2c^4 + p^2 c^2 \right)^{1/2}$ with $ \hat{p} = -i\hbar {\partial}/{\partial x}$. These results are consistent with those of \cite{adak2004_1}.

\subsection{An exact solution to the Dirac equation}

We make a spinor ansatz in order to obtain an exact solution to the Dirac equation (\ref{ref:Dirac_eq_like_Sch})
\begin{equation}
    \psi= \begin{pmatrix}
        \alpha(x)  \\
        \beta(x) 
   \end{pmatrix} e^{i\Omega t} \label{ansatz_psi}
\end{equation}
where $\alpha(x)$ and $\beta(x)$ are complex unknown functions and  and $\Omega$ is a real constant. Afterwards it gives rise to two scalar coupled differential equations
\begin{subequations}
\begin{align}
      f\alpha' + (N - \frac{mc}{\hbar})\alpha + \frac{i\Omega}{c}\frac{\beta}{f} &=0 \ , \label{Dirac_feq0} \\
      f\beta' + (N + \frac{mc}{\hbar})\beta + \frac{i\Omega}{c}\frac{\alpha}{f} &=0 \ . \label{Dirac_feq1}
\end{align}
\end{subequations}
We could not succeed to find a class of exact solutions to them. Therefore we omit ${mc}/{\hbar}$ term by assuming $N >> {mc}/{\hbar}$ and then could find a solution for $a_2 = 1$ meaning $N = C_0/2$ and $Q = 0$ as
\begin{subequations}
\begin{align}
    \alpha &= \frac{1}{\sqrt{C_1 + C_0 x}}\left( D_0\, e^{\displaystyle\frac{i\Omega}{c\, C_0(C_1 + C_0 x)}} + D_1\, e^{-\displaystyle\frac{i\Omega}{c\, C_0(C_1 + C_0 x)}} \right) , \\
    \beta &= \frac{1}{\sqrt{C_1 + C_0 x}}\left( D_0\, e^{\displaystyle\frac{i\Omega}{c\, C_0(C_1 + C_0 x)}} - D_1\, e^{-\displaystyle\frac{i\Omega}{c\, C_0(C_1 + C_0 x)}} \right)
\end{align}
\end{subequations}
where $D_0$ and $D_1$ are arbitrary constants. Here we checked whether this explicit solution to the Dirac equation (\ref{dirac equation}) satisfies $\tau_a[\psi]=0$ and $\Sigma_a{}^b[\psi]=0$, defined in the equations (\ref{psienergymomentum}) and (\ref{psiangularmomentum}), respectively, and observed that it does not. In brief, we have found a solution of the vacuum theory and then a solution of the Dirac equation in a specific background. As we remarked above, just after (\ref{C0 constant}), $C_0$ behaves like an implicit source of gravity and $N$ is proportional to $C_0$ via the equation (\ref{N definition}). Correspondingly $N$ should contain a kind of gravitating mass. By assuming $N >> {mc}/{\hbar}$ we think that the gravitating mass is much larger than the mass of Dirac particle. Both solutions have the character of a damping oscillator because both wave functions go to zero as $x \rightarrow \pm \infty$. We want to pay attention that both of the solutions have a singularity at the point $x_0=-C_1/C_0$. In fact this singularity belongs to the spacetime itself as we discussed just after the equation (\ref{singularity analysis}). So, our spinor solutions are not valid at that point. Physically we interpret it as that the Dirac particle (or its anti-particle) can not cross that point. This ought to be the picture because we are in one space dimension and there is a gravitating source in somewhere, probably at $x_0=-C_1/C_0$, on the line.        


\section{Conclusions}
We have studied a new gravity model written in terms of the nonmetricity and the full curvature tensors and the minimal coupling of a Dirac particle to it. Firstly we summarized some basic concepts for the exterior algebra which we used through all the paper. Then we put some effort on the issue of the group formed by the transformation elements of the general linear coordinate transformations. We arrive at the conclusion that the group of those transformations is the Lorentz group in the orthonormal coframe bundle independently from the existence of nonmetricity or not. It is the reason of naming the orthonormal coframe also the Lorentzian coframe. Afterwards, since we would adhere the Lagrange formulation for developing our theory, we gave an analysis leading us to determine the form of our Lagrangian. It is well known that the gauge theory is very successful at explaining the microscopic phenomena. Therefore, by making an analogy between the Maxwell-Dirac theory, which contains very basics of the gauge theory, and our new model of gravity we decide a Lagrangian quadratic in nonmetricity and curvature tensors. In our comprehension, the metric tensor represents the physical gravitational field and the full connection represents the gauge field/potential. In order to trace the effects of the general relativity we keep the non-Riemannian Einstein-Hilbert term in the Lagrangian. In the next step we introduced the exterior covariant derivative of the Dirac spinor and minimally coupled the Dirac Lagrangian to our gravitational Lagrangian. Meanwhile, up to this point all the discussions and results are valid in any dimensions. Now we obtained the variational field equations in two dimensions because the investigation  of general 2D gravity models allows 
to tackle fundamental questions about quantum gravity by overcoming   challenges and technical complications  of higher dimensions.

In addition, we searched some classes of solutions in order to close our investigation on a modified theory of gravity. We tried finding a stationary solution, but since we could not obtain an exact solution we decided following a simplifying strategy. Thus we omitted the energy momentum, $\tau_a[\psi]$, and angular momentum (or hypermomentum), $\Sigma_a[\psi]$, tensors of the Dirac field in the coframe and the connection equations. Correspondingly we could obtain several classes of vacuum solutions and noticed that some of them are the same of the reference \cite{adak2008}. In our main solution we have a peculiar constant given by (\ref{C0 constant}) that can be seen as a representative of a kind of gravitating object. After giving some discussions on these solutions we dealt with the Dirac equation representing a spinning particle in a specific background geometry with both nonmetricity and full curvature. Firstly we obtained the Hamiltonian matrix by comparing the Dirac equation to the Schrödinger equation and calculated its eigenvalues. We remarked that the results agreed with those of the reference \cite{adak2004_1}. Finally we could find an exact solution to the Dirac equation behaving like a damped oscillator under certain circumstances. Our future research plan is to improve this work by including more quadratic terms in both the curvature and the nonmetricity in the Lagrangian.

\section*{Acknowledgments}

We thank to the anonymous referees for their guiding questions and criticisms. In this study, M.A. was supported via the project number 2020KRM005-197 and O.S. via the project number 2020KRM005-010 by the Scientific Research Coordination Unit of Pamukkale University.

\section*{References}

\renewcommand{\section}[2]{} 

\end{document}